\documentclass[10pt,twocolumn,a4paper]{article}

\usepackage[left=20mm, right=15mm, top=15mm, bottom=15mm]{geometry}

\usepackage{subfig}
\usepackage{flushend}
\usepackage{indentfirst}
\usepackage{graphics}
\usepackage{amsmath}
\usepackage{graphicx}
\usepackage{epstopdf}
\usepackage{float}

\begin{document}

\title{\huge \textbf{On Exact Cosmological Models of a Scalar \\Field in Lyra Geometry
}}

\date{}

\twocolumn[
\begin{@twocolumnfalse}
\maketitle

\author{\textbf{V. K.  Shchigolev}$^{1,*}$\\\\
\footnotesize $^{1}${Department of Theoretical Physics, Ulyanovsk State University, 42 L. Tolstoy Str., Ulyanovsk 432000, Russia}\\

\footnotesize $^{*}$Corresponding Author: vkshch@yahoo.com}\\\\\\

\end{@twocolumnfalse}
]

\noindent \textbf{\large{Abstract}} \hspace{2pt} Exact cosmological models for a scalar field in Lyra geometry are studied in the presence of a time-varying effective cosmological term originated from the specific interaction of an auxiliary $ \Lambda $ -  term with the  displacement vector.  In this case, some exact solutions for the model equations are obtained with the help of the so-called superpotential method (or the first-order formalism). Some possible ways of further developing for such a model are offered.\\

\noindent \textbf{\large{Keywords}} \hspace{2pt} Cosmological Models, Scalar Field, Cosmic Evolution, Exact Solutions, Superpotential Function\\

\noindent\hrulefill

\section{\Large{Introduction}}

As well known, Lyra's geometry \cite{Lyra} can be considered as the candidate for modification of the contemporary cosmological models, the necessity of which is almost generally recognized. The great number of the modified theories of gravity is known, but now those modifications attract attention of researchers again due to the discovery of the late-time cosmological acceleration. The latter definitely followers from  the supernovae of type Ia observations \cite{Riess, Perlmutter}, Cosmic Microwave Background Radiation \cite{Spergel, Komatsu}, and Baryon Acoustic Oscillations in galaxy surveys \cite{Blake, Seo}. Variety of versions of the scalar-tensor theories are also involved in solving  the problem of cosmic acceleration. Among them, the most studied is the Brance - Dicke theory of gravitation \cite{Brans, Dicke}. However, some other theories gained increasing attention due to the same problem of cosmic acceleration. One example of such a theory is known as the theory of gravity in Lyra geometry.

Besides all, a substantial number of investigations of the models with a varying in time cosmological term has been proposed during the last two decades (see, e.g.  [10-18]). A number of interesting results have been obtained when accounting for the effective cosmological term in Lyra geometry [19-24].
Recall that the displacement vector field has a purely geometrical origin. Therefore, it seems rather natural that such a field in the dynamical equations of cosmological models should be considered as a constant associated in some way with a cosmological constant , or as a hidden parameter which is included  implicitly into the dynamical parameters of model. Namely this approach to the displacement vector is proposed in our work. Generally speaking, such an idea is not something absolutely new.  It is involved  in many studies of the models with variable $G$ and decaying vacuum energy density. Our aim is to apply this idea to the models in Lyra geometry.

The assumption of decaying cosmological term makes possible to preserve the continuity equation for the ordinary matter in its standard form, and generate a new (effective) cosmological term. Note that appearance of such a term is entirely due  to the influence of the displacement field.  It determines, along with matter, the dynamics of the cosmological evolution and becomes a constant in its absence, as it should be in the case of General Relativity.

In this paper,  the scalar-field model with an effective cosmological term in Lyra geometry is studied and discussed. Besides  the general consideration, we employ the so-called superpotential method in order to obtain some exact solutions of our model. We suggest some possible ways of further developing for such a model.

\section{\Large{A Brief Summary of Lyra Geometry}}

A scalar-tensor theory of gravitation proposed by Sen and Dunn \cite{Dunn} based on a Lyra manifold rather than a Riemannian manifold. It is shown that this new theory predicts the same effects, within observational limits, as in Einstein's theory.
The Lyra geometry \cite{Lyra} can be considered as a modification of the Riemann geometry in the direction similar to the Weil geometry. Here we present a brief summary of Lyra geometry which is necessary  for further study of our model.

In Lyra geometry, the displacement vector between two neighboring points $x^i$ and $x^i + d x^i $ is determined by the components $\Psi\,dx^i$, where $\Psi = \Psi (x^k)$ is a gauge function.
Then, the coordinate system $x^i$ and the gauge function $\Psi$ constitute a reference frame  $(\Psi, x^i)$. The transition to the new reference frame $(\Psi ', x'^i)$ is given by
\begin{equation}\label{1}
\Psi'=\Psi(\Psi, x^k),~~~x'^i=x^i(x^k),
\end{equation}
where $\partial \Psi'/\partial \Psi \ne 0,~~ det|\partial x'^i/\partial x^k| \ne 0 $.
The Levi-Civita connection coefficients in Lyra geometry are defined as follows:
\begin{equation}\label{2}
* \Gamma^i_{jk}= \Psi^{-1} \Gamma^i_{jk} - \frac{1}{2}(\delta^i_j \phi_k +\delta^i_k \phi_j - g_{jk}\phi^i),
\end{equation}
where $\Gamma ^i_{jk}$ is defined in terms of the metric tensor $g_{ik}$  just like as in Riemannian geometry, and $\phi_k$ is the displacement vector field. Lyra \cite{Lyra} and Sen \cite{Dunn} have showed that any general frame of reference vector field $\phi_k$ arises as a natural consequence of the introduction of the gauge function $\Psi$ in the structureless manifold. Note that $*\Gamma^i_{jk}$ is symmetric in the two lower indices.

The metric on the Lyra manifold is determined by the interval
\begin{equation}\label{3}
d s^2 = \Psi^2 g_{ik} d x^i d x^k,
\end{equation}
which is invariant with respect to the coordinate and  gauge transformations.

As a result, the parallel transport of a vector $\xi^i$ is given by
\begin{equation}\label{4}
d \xi ^i= - \tilde{\Gamma} ^i_{jk}\xi^j \Psi d x^k,
\end{equation}
where
\begin{equation}\label{5}
\tilde{\Gamma} ^i_{jk}= * \Gamma ^i_{jk}- \frac{1}{2}\delta^i_j \phi_k.
\end{equation}
Thus, we see that $\tilde {\Gamma}^i_{jk}$ is not symmetric with respect to $j$ and $k$. A remarkable difference from the Weyl geometry is that in Lyra geometry length of the vector does not change under parallel transport.

As always, the curvature tensor is defined in terms of the parallel transport of a vector along a closed curve, and is equal to
\begin{eqnarray}\label{6}
* R^i_{.jkl} = \Psi^{-2}[-(\Psi\tilde{\Gamma} ^i_{jk})_{,l}+ (\Psi\tilde{\Gamma} ^i_{jl})_{,k} \nonumber\\\nonumber\\- \Psi ^2 (\tilde{\Gamma} ^m_{jk}\tilde{\Gamma} ^i_{ml}-\tilde{\Gamma} ^m_{jl}\tilde{\Gamma} ^i_{km})]
\end{eqnarray}
where $\tilde{\Gamma} ^i_{jk}$ is determined from (\ref{5}). The convolution of the curvature tensor (\ref{6}) yields the scalar curvature
\begin{equation}\label{7}
* R = \Psi^{-2} R + 3 \Psi ^{-1} \phi^i_{;i}+ \frac{3}{2} \phi^i\phi_i
+ 2 \Psi^{-1}(\ln \Psi^2)_{,i} \phi^i,
\end{equation}
where $R$ is the Riemannian  scalar of curvature, and the semicolon denotes the covariant
derivative with the Christoffel symbols of the second kind.

The action integral is invariant  under the gauge and coordinate transformations , and is given in the form
\begin{equation}\label{8}
I= \int L \sqrt{-g}\Psi^4 d^4 x,
\end{equation}
where $d^4 x$ is a volume element, and $L$ is a scalar function.

Using the normal gauge  $\Psi = 1$ \cite{Sen}, and putting $L = * R$  in equation (\ref{8}), it is easy to find that equation (\ref{7}) can be reduced to the following form
\begin{equation}\label{9}
* R = R + 3 \phi^i_{;i}+\frac{3}{2}\phi^i \phi_i.
\end{equation}
The field equation is obtained from the variational principle
\begin{equation}\label{10}
\delta(I+J)=0,
\end{equation}
where $I$ is defined by (\ref{8}), and the action $J$ is defined by the Lagrangian density of matter ${\cal L}_m$ as usual:
\begin{equation}\label{11}
J = \int {\cal L}_m \sqrt{-g} d^4 x.
\end{equation}

\section{\Large{Gravitational Field Equations in Lyra Geometry}}

The Einstein's field equations in Lyra geometry in normal gauge and with a time-varying $\Lambda$ - term  can be written as
\begin{equation}\label{12}
R_{ik}- \frac{1}{2} g_{ik} R - \Lambda g_{ik} +  \frac{3}{2}\phi_i \phi_k - \frac{3}{4}g_{ik}\phi^j \phi_j = T_{ik},
\end{equation}
where $\phi_i$ is a displacement vector. For simplicity, we assume  that the gravity coupling constant $8\pi G=1$. All other symbols have their usual meanings in the Riemannian geometry.
The energy-momentum tensor (EMT) of matter $T_{ik}$ can be derived in a usual manner from the Lagrangian of matter (\ref{11}). Considering the matter as a perfect fluid, we have
\begin{equation}\label{13}
T_{ik}= (\rho_m +p_m)u_i u_k -p_m\, g_{ik},
\end{equation}
where  $u_i = (1,0,0,0)$ is  4-velocity of the co-moving observer, satisfying $u_i u^i = 1$.
Let $\phi_i$ be a time-like vector field of displacement,
\begin{equation}\label{14}
\phi_i = \left[\frac{2}{\sqrt{3}}\,\beta(t),0,0,0\right],
\end{equation}
where the numerical factor $2/\sqrt{3}$ is substituted for the sake of convenience in what follows. The line element of a spatially flat Friedmann-Robertson-Walker (FRW) space-time is represented by
$$
ds^2 = d t^2- a^2 (t)\Big( dx^2+dy^2+dz^2\Big),
$$
where $a(t)$ is a scale factor of the Universe.
Given this metric and equations (\ref{13}), (\ref{14}), we can reduce  (\ref{12}) to the following set of equations:
\begin{eqnarray}
3H^2  - \beta^2 &=& \rho_m +\Lambda,\label{15}\\
2 \dot H + 3H^2  + \beta^2 &=& -  p_m+\Lambda,\label{16}
\end{eqnarray}
where $H = \dot a/a $ is the Hubble parameter, and the overdot stands for  differentiation with respect to cosmic time $t$.

The continuity equation follows from (\ref{15}) and (\ref{16}) as:
\begin{equation}\label{17}
\dot \rho_m + \dot \Lambda + 2 \beta \dot \beta + 3 H \Big(\rho_m + p_m + 2\beta^2 \Big)=0.
\end{equation}

For more convenience, one can rearrange the basic equations of the model, (\ref{16}) and (\ref{17}), as follows:
\begin{equation}\label{18}
3H^2  = \rho_m +\Lambda + \beta^2,
\end{equation}
\begin{equation}\label{19}
2 \dot H  = - (\rho_m + p_m + 2  \beta^2).
\end{equation}
One could readily  verify that the continuity equation (\ref{17}) occurs from the set of equations (\ref{18}), (\ref{19}) in the form of equation . The reason of this is that the conservation equation for EMT $T^k_{i\,; k} = 0$, where semicolon stands for the covariant derivative, is a consequence of the identity  $G^k_{i\,; k} = 0$ for the Einstein tensor. We are going to preserve the continuity equation for matter in its standard form. Therefore, the field equation (\ref{12}) yields  the following equation for $\Lambda$ - term and displacement field:
$$
 \Lambda_{;i} =  \frac{3}{2}\Big(\phi_i \phi^k - \frac{1}{2}\delta_i^k\phi^j \phi_j\Big)_{;k}.
$$
For the homogeneous fields of $\Lambda(t)$ and $\beta(t)$, this equation becomes as follows
\begin{equation}\label{20}
\dot \Lambda + 2 \beta \dot \beta + 6 H \beta^2 =0.
\end{equation}
Taking into account this equation in  (\ref{19}), we can obtain the following continuity equation for matter:
\begin{equation}
\dot \rho_m + 3 H (\rho_m + p_m )=0.\label {21}
\end{equation}

Let us emphasize that  the displacement vector field has strongly geometric nature. We assume that it can give rise to an effective cosmological term $\Lambda_{eff}$. Therefore, we believe that $\Lambda(t)$ is not an independent dynamical parameter of the model, and it should be excluded from the system of equations (\ref{18}), (\ref{19}). At this, such a manipulation leads to an effective cosmological term via the kinematic (geometric) parameter of the model, namely $H(t)$, but only in the presence of the displacement vector field. Integrating (\ref{22}) for $\Lambda(t)$, we obtain
\begin{equation}\label{22}
\Lambda = -\beta^2-6\int H \beta^2 d t +\Lambda_0,
\end{equation}
where $\Lambda_0$ is a constant of integration.

Substituting (\ref{22}) into the basic set of equations (\ref{18}), (\ref{19}),  we are able to rewrite the main equations of our model as follows
\begin{eqnarray}
3H^2  &=& \rho_{eff},\label{23}\\
2 \dot H  &=& - (\rho_{eff} + p_{eff}),\label{24}
\end{eqnarray}
where we have introduced the effective cosmological term as
\begin{equation}\label{25}
\Lambda_{eff}(t)=\Lambda_0 - 6 \int H \beta^2 d t,
\end{equation}
and the effective energy density and pressure as
\begin{equation}\label{26}
\rho_{eff}=\rho_m +\Lambda_{eff},\,\,\,\,p_{eff}=p_m -\Lambda_{eff} - \frac{\dot\Lambda_{eff}}{3 H}.
\end{equation}

It can be verified that the effective energy density and pressure (\ref{26}) also satisfy the continuity equation in its usual form due to (\ref{23}), (\ref{24}) and the continuity equation for matter (\ref{21}):
\begin{equation}
\dot \rho_{eff} + 3 H (\rho_{eff} + p_{eff} )=0.\label {27}
\end{equation}

It is noteworthy that in our approach both matter and effective fluid  satisfy the continuity equation in the usual form. Therefore, we can introduce the equation of state (EoS) of matter $w_m = p_m/\rho_m$ and also a barotropic index of effective fluid $w_{eff}=p_{eff}/\rho_{eff}$ as
\begin{equation}
w_{eff} = -1-\frac{2 \dot H}{3 H^2}= -1 +\frac{\displaystyle w_m + 1 -\frac{\phantom{\dot A}\dot \Lambda _{eff}}{3 H \rho_m}}{\displaystyle 1+\frac{\phantom{\dot A}\Lambda_{eff}}{\rho_m}}\,.\label{28}
\end{equation}
As seen, the effective EoS may considerably deviate from EoS of matter. Even if $w_m$ is some constant, the effective EoS can vary in time. Even more remarkable feature of this model is the ability of $w_{eff}$ to cross the so-called phantom divide line $w_{eff}=-1$. As supposed, this crossing is to be highly probable according to the contemporary observational data [2-7].

\section{\Large{Scalar-Field Model}}

In this section, we consider a quintessence (or phantom) field as a source of gravity in the Universe. One can rewrite  equations (\ref{23}) and (\ref{24}) in terms of the effective energy density $\rho_{eff}$ and pressure $p_{eff}$, taking into account the following expressions for the scalar field $\varphi$:
\begin{equation}
\label{29} \rho_m = \epsilon\frac{\dot \varphi^2}{2} + V(\varphi),\quad p_m = \epsilon \frac{\dot \varphi^2}{2} - V(\varphi)~,
\end{equation}
where $\epsilon=+1$ represents quintessence while $\epsilon=-1$ refers to phantom field.
As a result, we have
\begin{eqnarray}
3H^2  &=& \epsilon\frac{\dot \varphi^2}{2} + V(\varphi)+\Lambda_{eff},\label{30}\\
2 \dot H  &=& - \epsilon \dot \varphi ^2+\frac{\dot\Lambda_{eff}}{3 H},\label{31}
\end{eqnarray}
so far as
$$
\rho_{eff}=\epsilon\frac{\dot \varphi^2}{2} + V(\varphi)+\Lambda_{eff},
$$
and
$$
p_{eff}= \epsilon \frac{\dot \varphi^2}{2} - V(\varphi)-\Lambda_{eff} - \frac{\dot\Lambda_{eff}}{3 H},
$$
in accordance with (\ref{26}) and (\ref{29}).
At last, in view of (\ref{25}) and (\ref{29}), the set of basic equations (\ref{30}), (\ref{31}) becomes
\begin{eqnarray}
3 H^2 &=& \epsilon\frac{\dot \varphi^2}{2} + V(\varphi)+\Lambda_0 -6 \int H \beta^2 d t, \label{32}\\
2 \dot H &=& - \epsilon\dot \varphi^2 -2\beta^2, \label{33}
\end{eqnarray}
that is, takes the form suitable for further study.

Here, we have to add the explicit expression for the field potential  obtained from equation  (\ref{32}) in the following form:
\begin{equation}\label{34}
V(\varphi)= 3 H^2+ 6 \int H \beta^2 d t -\epsilon\frac{\dot \varphi^2}{2}-\Lambda_0.
\end{equation}

\subsection{\normalsize \textbf{The Superpotential Method}}

One can make sure that it is hard to find exact solutions for this model.  Nevertheless, a wide class of exact solutions can be obtained in terms of the so-called superpotential. It is worth to note that this method was firstly developed for a single scalar field in \cite{Zhuravlev}. Later this method  has been re-opened again as the so-called first order formalism in \cite{Gomes}. As for applying this method to our model, first of all we have to redefine the geometrical field of displacement vector with the help of a new field $\alpha(t)$ as
\begin{equation}\label{35}
\beta^2(t)=\dot \alpha^2(t),
\end{equation}
where the overdot as usually stands for the derivative with respect to time. As a result, we have the following equation instead of (\ref{33})
\begin{equation}
2 \dot H = - \epsilon\dot \varphi^2 -2\dot \alpha^2(t), \label{36}
\end{equation}
Let us suppose that  the superpotential function $W(\varphi, \alpha)$ can be presented by the equation
\begin{equation}\label{37}
H=W(\varphi, \alpha),
\end{equation}
in which the Hubble parameter $H(t)$, as a function of time, is presumably expressed in terms of fields $\varphi(t),\, \,\alpha (t)$. Substituting (\ref{36})  into (\ref{37}), one can obtain two first-order equations as follows:
\begin{equation}\label{38}
\dot \varphi =-2 \epsilon W_{\varphi},\,\,\,\, \dot \alpha = -W_{\alpha} ,
\end{equation}
where and further  $W_{\varphi}\equiv\partial W/ \partial \varphi$ and $W_{\alpha}\equiv\partial W/ \partial \alpha$.
Taking into account (\ref{37}) and (\ref{38}), one can rewrite equation (\ref{25}) for the effective cosmological term as follows
\begin{equation}\label{39}
\Lambda_{eff}(t)=\Lambda_0 - 6 \int W W_{\alpha}^2 d t.
\end{equation}
The potential can be obtained from (\ref{34}) and (\ref{38}) in the following form:
\begin{equation}\label{40}
V(\varphi)= 3 W^2 + 6\int W W_{\alpha}^2 d\,t-2\epsilon W_{\varphi}^2 -\Lambda_0.
\end{equation}
As the potential depends on $\varphi$ alone, we should demand that $\partial V/\partial \alpha = 0$. Taking into account (\ref{38}) and (\ref{40}),  one can readily verify that this is equivalent to
\begin{equation}\label{41}
3 W W_{\alpha}=\epsilon W_{\varphi} W_{\varphi \alpha}.
\end{equation}
This equation can be satisfied in many ways depending on the structure of function $W(\varphi,\alpha)$. In the framework of our short paper, we consider two simplest examples of this function. For this end, let us introduce two differentiable functions, $X(\varphi)$ and $Y(\alpha)$, of the separated variables $\varphi$ and $\alpha$. With the help of these functions, we can represent the superpotential in a simple manner, as shown below. Two different versions of the superpotential with separated variables  are used as the simplest  examples to elucidate the solution procedure.

\subsubsection{\normalsize{The First Ansatz}}

Let us suppose that
\begin{equation}\label{42}
W(\varphi,\alpha) =X(\varphi)+Y(\alpha).
\end{equation}
Consequently, we have $ W_{\varphi}= X ' (\varphi)$ and $ W_{\alpha}= Y ' (\alpha)$.
In this case, the set  of equations (\ref{40}) becomes as follows
\begin{equation}\label{43}
\dot \varphi =-2 \epsilon X ' (\varphi),\,\,\,\, \dot \alpha = -Y ' (\alpha),
\end{equation}
that means the separation of variables. In addition, we should consider a restriction on these equations, arising from the equation (\ref{41}). Substituting (\ref{42}) into (\ref{41}), we obtain
$$
W(\varphi, \alpha)Y ' (\alpha) = 0.
$$

If we  still consider that the fields $\varphi, \alpha$ do not interact via the potential, that is this restriction  is valid and $H=W \neq 0$, then we arrive at $Y ' (\alpha) = 0$. Therefore, we have to conclude that this case is a trivial one. Indeed, we have $\beta^2=\dot \alpha^2 = Y ' (\alpha) = 0$ according to (\ref{43}). It means that we have this is the case of vanishing displacement vector. Due to (\ref{39}), the effective cosmological term becomes constant: $\Lambda_{eff}(t)=\Lambda_0$. As a result, we get a typical $\Lambda$CDM model. Such a trivial result encourages us to consider another ansatz.

\subsubsection{\normalsize{The Second Ansatz}}

Now we suppose that the superpotential is presented by
\begin{equation}\label{44}
W(\varphi, \alpha)=X(\varphi)\, Y(\alpha).
\end{equation}
that allows to rewrite the set of equations (\ref{38}) as
\begin{equation}\label{45}
\dot \varphi =-2 \epsilon X ' (\varphi) Y(\alpha),\,\,\,\, \dot \alpha = - X(\varphi) Y ' (\alpha),
\end{equation}
and the restriction equation (\ref{38}) - as
\begin{equation}\label{46}
Y(\alpha) Y '(\alpha) \Big[\epsilon X '\phantom{.}^2(\varphi) -3 X ^2(\varphi)\Big] = 0.
\end{equation}

We should decline the trivial case,  $Y(\alpha)Y '(\alpha)=0$, mentioned above and conclude that
\begin{equation}\label{47}
\epsilon X '\phantom{.}^2(\varphi) = 3 X ^2(\varphi).
\end{equation}
Staying in the field of the real functions, we have to study the case of $\epsilon = +1$.  Integrating equation (\ref{47}), we obtain
\begin{equation}\label{48}
X(\varphi)=-\lambda \exp[\pm \sqrt{3} (\varphi-\varphi_0)],
\end{equation}
where $\lambda>0$ and $\varphi_0$ are constants. In view of this function and  (\ref{45}), one can write down that
\begin{eqnarray}\label{49}
\dot \varphi &=&\pm 2\sqrt{3}\, \lambda \,e^{\displaystyle \pm \sqrt{3} (\varphi-\varphi_0)}\, Y(\alpha),\\
\dot \alpha &=&  \lambda\, e^{\displaystyle \pm \sqrt{3}(\varphi-\varphi_0)}\, Y '(\alpha).\label{50}
\end{eqnarray}
Making use of (\ref{49}) and (\ref{50}) in (\ref{36}), we yield
\begin{equation}\label{51}
\dot H = -\lambda^2 \,e^{\displaystyle \pm 2\sqrt{3} (\varphi-\varphi_0)}\,[ Y '^2(\alpha)+6\,   Y^2(\alpha)],
\end{equation}
which is not an independent equation. It can be derived from (\ref{44}). Indeed, inserting (\ref{49}) into (\ref{50}), we have
\begin{equation}\label{52}
H = W = -\lambda \,e^{\displaystyle \pm \sqrt{3}(\varphi-\varphi_0)}\, Y(\alpha).
\end{equation}
By differentiation of the latter with respect to time and taking into account (\ref{48}), (\ref{49}), we arrive at (\ref{51}).

It should be noted that function $Y(\alpha)$ is arbitrary one so far. To proceed further in solving the problem, we need to specify this function in some explicit form.  By doing so, one may follow some special restrictions on $Y(\alpha)$.  For example, this could be the requirement of solvability of the set of equations (\ref{49}), (\ref{50}). It is easy to note that combining these equations we arrive at the following one with the separated variables
\begin{equation}\label{53}
\dot \varphi = \pm 2\sqrt{3}\frac{Y(\alpha)}{Y '(\alpha)} \dot \alpha.
\end{equation}
It can be readily integrated if $Y/Y ' = Z '$, where $Z(\alpha)$ is some differentiable function.
The variety of functions $Y(\alpha)$ can meets this condition, such, for example,  as the power-law, exponential, trigonometric, hyperbolic functions and their combinations.

Let us demonstrate this approach, investigating a specific example. Consider the case of $Y(\alpha) = \sinh(C \alpha)$, where $C$ is a constant. Hence, we have $Y '(\alpha) = C \cosh(C \alpha)$. Combining this together with (\ref{49}) and (\ref{50}) , we can obtain the following equations
\begin{equation}\label{54}
\varphi=\pm \frac{2\sqrt{3}}{C^2}\,\ln\Big[\cosh(C \alpha)\Big] +\varphi_0,
\end{equation}
where $\varphi_0$ is a constant, and
\begin{equation}\label{55}
\dot \alpha =  \lambda C \left[\cosh(C \alpha)\right]^{\displaystyle 1+ 6 C^{-2}},
\end{equation}
Note that one can solve the last equation explicitly for any  integer power $(1+6C^{-2})>1$  in the right-hand-side of this equation. For simplicity, we consider only the case $C^2=6$. Integrating equation  (\ref{55}), we can find that
\begin{equation}\label{56}
C \alpha = \frac{1}{2} \,\ln\left[\frac{1+6\lambda(t-t_0)}{1-6\lambda(t-t_0)}\right],
\end{equation}
where $t_0$ is a constant.
Substituting (\ref{56}) into (\ref{52})-(\ref{55}), we have
\begin{eqnarray}
\label{57}  H(t) &=& -\frac{6\lambda^2(t-t_0)}{1-36 \lambda^2 (t-t_0)^2}, \\
\label{58}  \varphi(t) &=& \mp \frac{1}{2\sqrt{3}}\ln[1-36\lambda^2(t-t_0)^2]+\varphi_0, \\
\label{59}  \beta^2(t) &=& \dot \alpha^2 = 6\lambda^2 [1-36\lambda^2(t-t_0)^2]^{-1}.
\end{eqnarray}
By using these results in equation (\ref{40}), one can readily write down an explicit expression for the potential in the following form
\begin{equation}\label{60}
V(\varphi)=3 \lambda^2 \,e^{\displaystyle \pm \sqrt{3} (\varphi-\varphi_0)}\Big[2-\,e^{\displaystyle \pm \sqrt{3} (\varphi-\varphi_0)}\Big],
\end{equation}
where the value of additive constant is taken to be zero. Thus we have completed the reconstruction of the scalar field.

\section{\Large{Discussion and Conclusion}}

Let us briefly discuss the physical properties of the model. Equations (\ref{57})-(\ref{60}) represents a FRW universe in Lyra geometry which may be physically significant for the discussion of some stages of evolution of the universe. The different physical and kinematical parameters, which could be important to the physics of the model, are given below.

Integrating equation (\ref{57}), we can find that the scale factor is given by
\begin{equation}\label{61}
a(t) = a_0 \Big[1-36 \lambda^2 (t-t_0)^2\Big]^{1/12}.
\end{equation}
From equation (\ref{39}), one can calculate the effective cosmological term as
\begin{equation}\label{62}
\Lambda_{eff}=\Lambda_0 + \frac{3\lambda^2}{1-36 \lambda^2 (t-t_0)^2}.
\end{equation}

Then we can  obtain such an important physical parameter of the model as EoS. With the
help of (\ref{29}), (\ref{58}) and (\ref{60}), we can derive  that
\begin{equation}
w_m = 3-\frac{4}{72\lambda^2(t-t_0)^2+1}\,,\label{63}
\end{equation}
where $\Lambda_0$ considered being zero for the sake of simplicity. After that, the effective EoS can be found from equation (\ref{28}). Interesting to note that according (\ref{63}),  the EoS parameter of the scalar field achieves the quasi-vacuum  EoS at instant $t=t_0$ but never crosses it.

One of the most important  kinematical parameter of any cosmological model is the so-called deceleration parameter $q$. Using equation (\ref{57}) and the definition of $q$, we readily obtain
$$
q = -1 - \frac{\dot H}{H^2} = -1 + \frac{1+36\lambda^2(t-t_0)^2}{6 \lambda^2(t-t_0)^2}.
$$
From this equation, one can conclude that the model never expands with acceleration. Such a result is not in contradiction with (\ref{63}) because the evolution is determined by the effective EoS,  rather than the EoS of matter.

The  method of superpotential considered above of course not the only approach to generate exact solutions in such a model.
A wide class of solution for our model could
be obtained from the phenomenological laws for the evolution of cosmological term.  At this, we are able to consider many different laws for a time-varying cosmological term, represented in the literature \cite{Overduin}, \cite{Sahni}. Indeed, substituting the effective cosmological term   of the form $\Lambda_{eff}(t) = L(t,a,H)$, where $L(t,a,H)$ is a differentiable function, into (\ref{25}), we obtain after differentiation with respect to time \cite{Shchigolev3}:
$$
\beta^2 = -\frac{1}{6H}\frac{\partial L}{\partial t} -\frac{1}{6H}\frac{\partial L}{\partial a} \dot a - \frac{\dot H}{6H}\frac{\partial L}{\partial H}.
$$
This equation can be considered as the main one for searching  $a(t)\Rightarrow \beta(t)$ or $\beta(t)\Rightarrow a(t)$. After that,  the rest parameters of the model can be obtained from (\ref{32})-(\ref{34}).

Thus, we have studied and briefly discussed the scalar-field model with the effective cosmological term in Lyra geometry. In addition to the general results obtained in this article, we have considered one example of solving our model explicitly. It is worth to emphasize that this model is the analytically accurate and, therefore, suitable for the deeper study of the main features of cosmic evolution. We express the hope for further development of our model in terms with the help of some other methods.

\noindent\hrulefill


\begin{thebibliography}{99}

\small {

\bibitem{Lyra} G. Lyra, Sber eine Modifikation der Riemannschen Geometrie, Mathematische Zeitschrift, vol. 54, pp. 52-64, 1951.

\bibitem{Riess} A. G. Riess, et al.,  Observational Evidence from Supernovae for an Accelerating Universe and a Cosmological Constant,  Astron. J., Vol. 116, 1009, 1998.

\bibitem{Perlmutter} S. Perlmutter, et al.,  Measurements of Omega and Lambda from 42 High-Redshift Supernovae,  Astrophys. J.,  Vol. 517,  565, 1999.

\bibitem{Spergel} D.\,N. Spergel, et al., First-Year Wilkinson Microwave Anisotropy Probe (WMAP) Observations: Determination of Cosmological Parameters, Astrophys. J. Suppl. Ser., Vol. 148, 175-194, 2003.

\bibitem{Komatsu} E. Komatsu,  et al., Seven-Year Wilkinson Microwave Anisotropy Probe (WMAP) Observations: Cosmological Interpretation, Astrophys. J.S., Vol. 192, 18, 2011.

\bibitem{Blake} C. Blake , K. Glazebrook, Probing Dark Energy Using Baryonic Oscillations in the Galaxy Power Spectrum as a Cosmological Ruler, Astrophys. J., Vol. 594, 665, 2003.

\bibitem{Seo} H.-J. Seo  and D.J. Eisenstein, Improved Forecasts for the Baryon Acoustic Oscillations and Cosmological Distance Scale,  Astrophys. J., Vol. 665, 14-24, 2007.

\bibitem{Brans} C. Brans and R.H. Dicke, Mach's Principle and a Relativistic Theory of Gravitation, Phys. Rev., Vol. 124, 925, 1961.

\bibitem{Dicke} R.H. Dicke, Mach's Principle And Invariance Under Transformation Of Units, Phys. Rev., Vol. 125, 2163, 1962.

\bibitem{Chen} W. Chen and Y.-S. Wu, Implications of a cosmological constant varying as $R^{-2}$,   Phys. Rev. D, Vol. 41,  695, 1990.

\bibitem{Pavon} D. Pavon,  Nonequilibrium fluctuations in cosmic vacuum decay, Phys. Rev. D, Vol. 43,  375, 1991.

\bibitem{Carvalho} J. C. Carvalho , J. A. S. Lima  and  I. Waga, Cosmological consequences of a time-dependent $\Lambda$ term, Phys. Rev. D, Vol. 46, 2404, 1992.

\bibitem{Arbab} A. I. Arbab  and  A.-M. M. Abdel-Rahaman, Nonsingular cosmology with a time-dependent cosmological term, Phys. Rev. D, Vol. 50,  7725, 1994.

\bibitem{Matyjasek} J. Matyjasek, Cosmological models with a time-dependent $\Lambda$ term, Phys.Rev. D, Vol. 51,  4154, 1995.

\bibitem{Overduin} J. M. Overduin ,  F. I. Cooperstock, Evolution of the scale factor with a variable cosmological term, Phys. Rev. D, Vol. 58, 043506, 1998.

\bibitem{Sahni}  V. Sahni, A. Starobinsky, The Case for a Positive Cosmological $\Lambda$-Term, Int. J. Mod. Phys. D, Vol. 9, 373, 2000.

\bibitem{Carneiro} S. Carneiro , J. A. S. Lima, Time Dependent Cosmological Term and Holography, Int. J. Mod. Phys. A, Vol. 20, 2465, 2005.

\bibitem{Darabi} F. Darabi, Time variation of $G$ and $\Lambda$, acceleration of the universe, coincidence problem and Mach's cosmological coincidence,  Online available from http://arxiv.org/abs/0802.0028

\bibitem{Hova} H. Hova, A dark energy model in Lyra manifold, J. Geom. Phys., Vol. 64, 146-154, 2013.

\bibitem{Zhi} Haizhao Zhi, Mengjiao Shi, Xinhe Meng,  An effective cosmological term from a new global 1-form in Lyra geometric cosmos model, Online available from http://arxiv.org/abs/1210.6431\\

\bibitem{Chaubey} R. Chaubey, A. K. Shukla, A New Class of Bianchi Cosmological Models in Lyra's Geometry, Int. J. Theor. Phys., Vol. 52, 735, 2013.

\bibitem{Sahu}  S. K. Sahu , Tapas  Kumar,  Tilted Bianchi Type-I Cosmological Model in Lyra Geometry, Int. J. Theor. Phys., Vol. 52, 793, 2013.

\bibitem{Shchigolev1} V.\,K. Shchigolev, E.\,A. Semenova, Scalar Field Cosmology in Lyra's Geometry,  Online available from http://arxiv.org/abs/1203.0917

\bibitem{Shchigolev2} V.\,K. Shchigolev,  Cosmological Models with a Varying $\Lambda$ -Term in Lyra's Geometry, Mod. Phys. Lett. A, Vol. 27, No. 29, 1250164, 2012.

\bibitem{Dunn}	D. K. Sen and K. A. Dunn,  A scalar-tensor theory of gravitation in a modified Riemannian manifold, Journal of Mathematical Physics, Vol. 12, pp. 578-586, 1971.

\bibitem{Sen} D. K. Sen, A static cosmological model, Z. Phys. C, Vol. 149, pp. 311-323, 1957.

\bibitem{Zhuravlev} V. M. Zhuravlev, S. V. Chervon, Cosmological Inflation Models Admitting Natural Emergence to the Radiation-Dominated Stage and the Matter Domination Era, Zh. Eksp.Teor. Fiz., Vol. 91, No. 2, pp. 227–238, 2000.

\bibitem{Gomes} D. Bazeia, C. B. Gomes, L. Losano, R. Menezes, First-order formalism and dark energy, Phys. Lett. B, Vol. 633, 415, 2006.

\bibitem{Shchigolev3} V.\,K. Shchigolev,  Cosmology with an Effective $\Lambda$-Term in Lyra Manifold, Online available from http://arxiv.org/abs/1307.1866
}

\end{thebibliography}
\end{document}